# A Cost Effective RFID Based Customized DVD-ROM to Thwart Software Piracy

Prof. Sudip Dogra
Electronics & Communication Engineering
Meghnad Saha Institute of Technology
Kolkata, India

Prof. Subir Kr. Sarkar
Electronics and Telecommunication Engineering
Jadavpur University
Kolkata, India

Ritwik Ray
Student
Electronics & Communication Engineering
Meghnad Saha Institute of Technology
Kolkata, India

Saustav Ghosh
Student
Electronics & Communication Engineering
Meghnad Saha Institute of Technology
Kolkata, India

Debharshi Bhattacharya
Student
Electronics & Communication Engineering
Meghnad Saha Institute of Technology
Kolkata, India

*Abstract*—Software piracy has been a very perilous adversary of the software-based industry, from the very beginning of the development of the latter into a significant business. There has been no developed foolproof system that has been developed to appropriately tackle this vile issue. We have in our scheme tried to develop a way to embark upon this problem using a very recently developed technology of RFID.

## I. INTRODUCTION

OVER the years, the software industry has developed into a multi-billion dollars business, with it spreading its wings throughout the world. Not only in the commercial field, but softwares are now being applied in almost all spheres of our life. Ranging from defense activities to health monitoring, there are softwares for every purpose. As a result, these softwares come with varying price tags. Softwares used in scholarly, medical or defense activities are generally highly priced because of their significance. The utmost peril that has been menacing this exceptionally vital industry is the act of software piracy.

In our present work, we have tried to develop a DVD-ROM which will be capable of reading only the authorized DVDs containing softwares, and will be used only for the purpose of storing costly sensitive data. For this purpose, we have taken the help of the latest RFID technology. We have discussed about RFID and the functioning of a DVD-ROM in sections II and III respectively. Following which, a brief discussion about software piracy has been done in sections IV and V. After this, we have described our scheme and listed the advantages in sections VI and VII respectively.

## II. RFID: RADIO FREQUENCY IDENTIFICATION

RFID stands for Radio Frequency IDentification, a term that describes any system of identification wherein an electronic device that uses radio frequency or magnetic field variations to communicate is attached to an item. The two most talked-about components of an RFID system are the tag, which is the identification device attached to the item we want to track, and the reader, which is a device that can recognize the presence of RFID tags and read the information stored on them. The reader can then inform another system about the presence of the tagged items. The system with which the reader communicates usually runs software that stands between readers and applications. This software is called RFID middleware.

In a typical RFID system [2], passive tags are attached to an object such as goods, vehicles, humans, animals, and shipments, while a vertical/circular polarization antenna is connected to the RFID reader. The RFID reader and tag can radio-communicate with each other using a number of different frequencies, and currently most RFID systems use unlicensed spectrum. The common frequencies used are low frequency (125 KHz), high frequency (13.56 MHz), ultra high frequency (860–960 MHz), and microwave frequency (2.4 GHz). The typical RFID readers are able to read (or detect) the tags of only a single frequency but multimode readers are becoming cheaper and popular which are capable of reading the tags of different frequencies [3].

### III. OPERATION OF A DVD-ROM

A DVD ROM is very similar to a CD ROM. It has a laser assembly that shines the laser beam onto the surface of the disc to read the pattern of bumps. The DVD player decodes the encoded Data, turning it into a standard composite digital signal. The DVD player has the job of finding and reading the data stored as bumps on the DVD. Considering how small the bumps are, the DVD player has to be an exceptionally precise piece of equipment. The drive consists of three fundamental components:

- A drive motor to spin the disc - The drive motor is precisely controlled to rotate between 200 and 500 rpm, depending on which track is being read.

- A laser and a lens system to focus in on the bumps and read them - The light from this laser has a smaller wavelength (640 nanometers) than the light from the laser in a CD player (780 nanometers), which allows the DVD laser to focus on the smaller DVD pits.

- A tracking mechanism that can move the laser assembly so the laser beam can follow the spiral track - The tracking system has to be able to move the laser at micron resolutions.

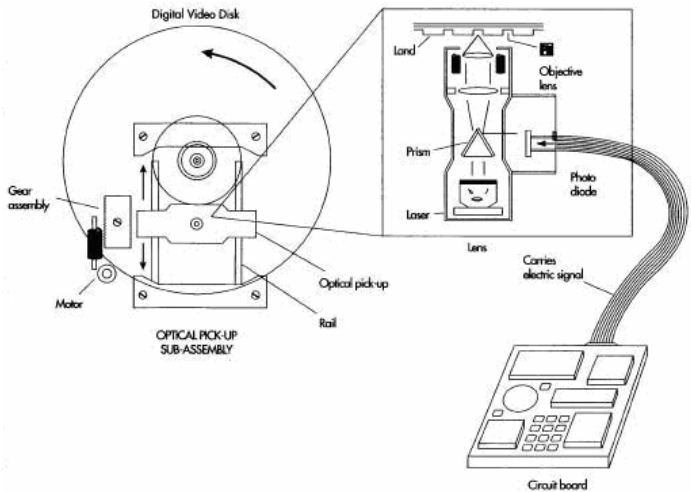

Fig. 1. Functional Diagram of a DVD-ROM

Inside the DVD player, there is a good bit of computer technology involved in forming the data into understandable data blocks, and sending them either to the DAC, in the case of audio or video data, or directly to another component in digital format, in the case of digital video or data. The fundamental job of the DVD player is to focus the laser on the track of bumps.

The laser can focus either on the semi-transparent reflective material behind the closest layer, or, in the case of a double-layer disc, through this layer and onto the reflective material behind the inner layer. The laser beam passes through the polycarbonate layer, bounces off the reflective layer behind it and hits an opto-electronic device, which detects changes in light. The bumps reflect light differently than the "lands," the flat areas of the disc, and the opto-electronic sensor detects that change in reflectivity. The electronics in the drive interpret the changes in reflectivity in order to read the bits that make up the bytes.

### IV. SOFTWARE PIRACY: A MODERN MENACE

Over the years, the software industry has developed into a multi-billion dollars business, with it spreading its wings throughout the world. Not only in the commercial field, but softwares are now being applied in almost all spheres of our life. Ranging from defense activities to health monitoring, there are softwares for every purpose. As a result, these softwares come with varying price tags. Softwares used in scholarly, medical or defense activities are generally highly priced because of their significance. The utmost peril that has been menacing this exceptionally vital industry is the act of software piracy.

The copyright infringement of software (often referred to as software piracy) refers to several practices which involve the unauthorized copying of computer software. Copyright infringement of this kind is extremely common. Most countries have copyright laws which apply to software, but degree of enforcement varies. After a dispute over membership between Iran and USA led to the legalization in Iran of the unconstrained distribution of software (see Iran and copyright issues), there have been fears that world governments might use copyright politically.

When software is pirated, customers, software developers, and resellers are harmed. Software piracy increases the risk consumer's computers will be corrupted by malfunctioning software and infected with viruses. Those who supply defective and illegal software do not tend to provide sales and technical support. Pirated software usually has insufficient documentation, which prevents consumers from enjoying the full benefits of the software package. In addition, consumers are not capable to take advantage of technical support and product upgrades, which are typically available to legitimate registered users of the software. Pirated software can cost consumers lost time and additional money.

| | | 2007 | 2006 | 2005 | 2004 | 2003 |
|---|---|---|---|---|---|---|
| | **ASIA-PACIFIC** | | | | | |
| | Australia | 28% | 29% | 31% | 32% | 31% |
| | Bangladesh | 92% | 92% | – | – | – |
| | China | 82% | 82% | 86% | 90% | 92% |
| | Hong Kong | 51% | 53% | 54% | 52% | 52% |
| | India | 69% | 71% | 72% | 74% | 73% |
| | Indonesia | 84% | 85% | 87% | 87% | 88% |
| | Japan | 23% | 25% | 28% | 28% | 29% |
| | Malaysia | 59% | 60% | 60% | 61% | 63% |
| | New Zealand | 22% | 22% | 23% | 23% | 23% |
| | Pakistan | 84% | 86% | 86% | 82% | 83% |
| | Philippines | 69% | 71% | 71% | 71% | 72% |
| | Singapore | 37% | 39% | 40% | 42% | 43% |
| | South Korea | 43% | 45% | 46% | 46% | 48% |
| | Sri Lanka | 90% | 90% | – | – | – |
| | Taiwan | 40% | 41% | 43% | 43% | 43% |
| | Thailand | 78% | 80% | 80% | 79% | 80% |
| | Vietnam | 85% | 88% | 90% | 92% | 92% |
| | Other AP | 91% | 86% | 82% | 76% | 76% |
| | **TOTAL AP** | **59%** | **55%** | **54%** | **53%** | **53%** |

Fig.2. Rate of Software Piracy across the countries (Cortesy:IDC)

Developers lose revenue from pirated software, from current products as well as from future programs. When software is sold most developers invest a portion of the revenue into future development and superior software packages. When software is pirated, software developers lose revenue from the sale of their products, which hinders development of new software and stifles the growth of the software company.

### V. SOFTWARE PIRACY: TYPES AND PREVENTIVE MEASURES

There are numerous kinds of software piracy. The bottom line is once software is pirated, the developer does not receive reparation for their toil. We have mentioned a few methods, which have been used contemporarily to check this despicable practice

#### A. End User Piracy

Using multiple copies of a single software package on several different systems or distributing registered or licensed copies of software to others. Another common form of end user piracy is when a cracked version of the software is used. Hacking into the software and disabling the copy protection or illegally generating key codes that unlocks the trial version making the software a registered version creates a cracked version.

#### B. Reseller Piracy

Reseller piracy occurs when an unscrupulous reseller distributes multiple copies of a single software package to different customers this includes preloading systems with software without providing original manuals & diskettes. Reseller piracy also occurs when resellers knowingly sell counterfeit versions of software to unsuspecting customers.

#### C. Trademark/Trade Name Infringement

Infringement occurs when an individual or dealer claims to be authorized either as a technician, support provider or reseller, or is improperly using a trademark or trade name. . Indications of reseller piracy are multiple users with the same serial number, lack of original documentation or an incomplete set, and non-matching documentation.

#### D. BBS/Internet Piracy

BBS/ Internet Piracy occur when there is an electronic transfer of copyrighted software. If system operators and/or users upload or download copyrighted software and materials onto or from bulletin boards or the Internet for others to copy and use without the proper license. Often hackers will distribute or sell the hacked software or cracked keys. The developer does not receive any money for the software the hacker distributed. This is an infringement on the developer's copyright.

Another technique used by software pirates is to illegally obtain a registered copy of software. Pirates acquire the software once and use it on multiple computers. Purchasing software with a stolen credit card is another form of software piracy.

Usually, the softwares are sold in the market in secondary memory devices like CDs and DVDs. Necessary measures are taken so that, the disks are copy protected and there are no likelihood of replicating the valuable software stored in it. Table I enlists some of the present technologies available for this purpose.

TABLE I
**Various existing Technologies used in the prevention of Software piracy**

| Serial No. | Name. | Description |
|---|---|---|
| 1. | Alkatraz | Copy protection for CD and DVD based on a "watermark" system |
| 2. | CD-Cops | CD-Cops is a envelope protection which is added to the CD's main executable. |
| 3. | CDShield | CDSHiELD protect your CD (before burning it) with putting voluntary sectors-errors to prevent copying from third unauthorized persons. |
| 4. | HexaLock | HexaLock CD-RX media are specially made CD-R's that contain a pre-compiled session, which includes security elements that make the discs copy protectable. |
| 5. | Laser Lock | LaserLock uses a combination of encryption software and a unique laser marking a "physical signature" on the CD surface made during the special LaserLock glass mastering procedure, in order to make copying virtually impossible. |
| 6. | Roxxe | Roxxe CD protection is a brand new combination of hardware and software protection that makes it impossible to run software from illegally copied CDs. |
| 7. | SafeDisc | Software publishers and developers need an effective and comprehensive anti-piracy solution to protect their intellectual property from copying, hacking and Internet distribution, while still ensuring a high quality experience for consumers. |
| 8. | SmarteCD | Smarte Solutions ("Smarte") is the leading provider of next generation Piracy Management solutions that secure and control the use of software and digital information while enhancing the |
| | | distribution and marketing-related capabilities of those products. |
| 9. | StarForce | StarForce Technologies is well known to the games and software world for its outstanding and hacker-proof copy protection systems for applications distributed on CD, DVD and CD-R. |

## VI. DESCRIPTION OF OUR PROPOSED SCHEME

### A. Our Consideration

In our scheme, we have proposed a modified DVD drive, in which only modified DVDs can be read. The Basic architecture of both the devices has been kept nearly the same. Only we have changed the working of the devices. The List of items used for our scheme is given in Table II.

TABLE II
**Components used in our scheme**

| Serial No. | Name. | Number |
|---|---|---|
| 1. | DVD-ROM | 1 |
| 2. | short range RFID reader | 1 |
| 3. | RFID passive tag | 4 |
| 4. | Computer | 1 |
| 5. | DVD | 4 |
| 6. | Basic Stamp Microcontroller | 1 |

Each of the DVDs will be fitted with a RFID Tag on the non-readable surface The Reader will be connected with the DVD-ROM. The interfacing will be done using a Basic Stamp Microcontroller. The power supply will provide the necessary power to run the reader, microcontroller and the DVD-ROM at the same time.

### B. Functioning of our Scheme

The basic principle underlying the mechanism of this scheme is that of authentication of two parties before the transfer of information actually begins. In our case the authentication process is carried out using the RFID technology. Each of the DVDs will be provided with a set of two serial numbers. One will be written on the DVD which will be visible to the user. The second code will be stored inside that of the RFID tag and can be read only by the reader. This code will have to be stored in a database inside the computer. If the process is carried out by a software company, then the second code will be given out in the internet in an encrypted form along with the serial number written on the DVD. The user will have to get this code first before he can run the DVD. A schematic diagram of the arrangement has been shown in Fig.3.

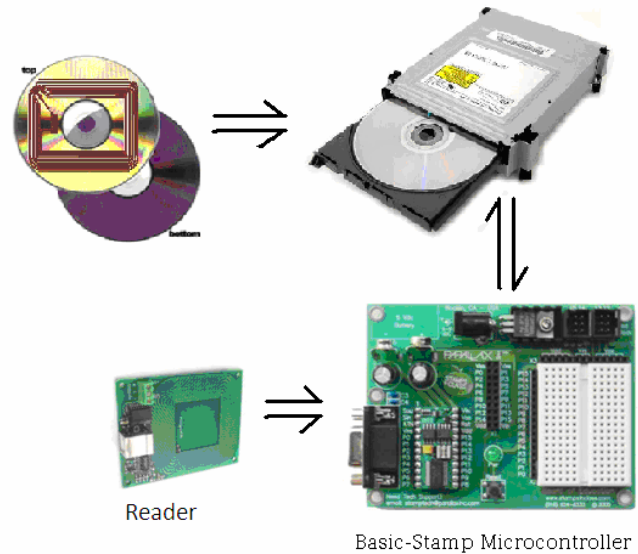

Fig. 3. Schematic Diagram of our arrangement

When this DVD is inserted into the drive the reader antenna will first read the code stored in the tag and send it to the microcontroller. Here the microcontroller will match this code with the ones existing in the computer's database. If the code does not match any of the previously existing codes, it will eject the DVD, and no data transfer will take place. It will send the signal to run the DVD only if it finds a match. Hence, the DVD-ROM won't be able to read no other DVDs other than the one having the authenticated RFID tags. The flowchart of the working has been shown in Fig.4.

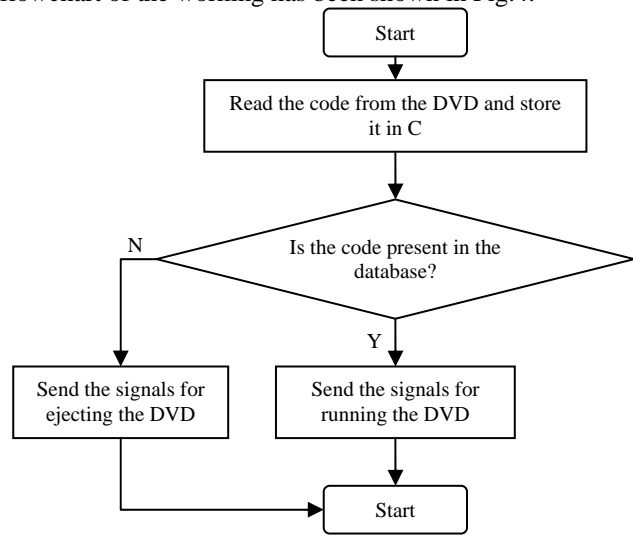

Fig. 4. Flowchart of the working of our scheme

Moreover, the DVD will be suitably encrypted so that it cannot be run on any other DVD-ROM, as well as the material stored in it wont be copied even by the modified DVD-ROM. We have simulated the signals that would be sent by the microprocessor using Verilog HDL, in Micro Sim. The simulations are showed in Fig. 5.

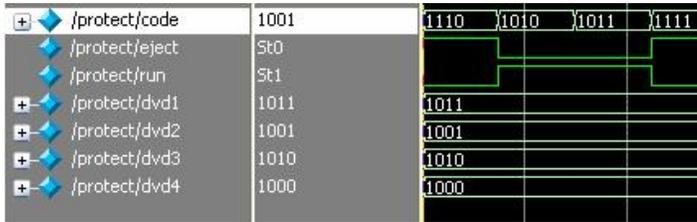

Fig. 5. Simulation of the various signals, used in our scheme, made in Verilog

As shown in the simulation, the codes stored in the 4 DVDs were 1000, 1001, 1010 and 1011. Whenever, a DVD having a false code is encountered, the Eject signal turns high, whereas the run signal turns low, making the DVD-ROM to eject the DVD. However, when the code matches with those of the database, the run signal turns high, and the eject signal goes low.

VII. ADVANTAGES OF OUR SCHEME

Over the years the piracy rackets in the Software industry has taken a huge toll in the section of losses incurred in the selling of this software. Numerous costly softwares like Operating System, Antivirus, etc are available in cheap CD/DVDs in the illegal markets in many parts of the world. Our scheme offers a cost effective solution in tackling this problem. The following advantages can be easily pointed out.

1) Since the special RFID DVD can only be run using a RFID optical drive, there will be very little possibility of the content being copied, as the DVD wont start running without proper authentication .

2) As the DVD will be completely made especially for the purpose of selling costly software, there will be proper configuration of the hardware, so that there will be neither any chance of  transferring the software data into any computer nor any chance of ripping the DVD.

3) New and advanced software are being launched everyday, which will eventually take the place of the older ones in the market. If our scheme is implemented by the software based companies, it will prevent the newer versions of the existing software to be available cheaply through piracy. Hence, the customer using the older version will be forced to buy the newer version only from the sources selling the original versions.

4) In view of the decreasing prices of the RFID readers and tags, a cheaper version of the modified DVD and its reader will be easily realizable for the customers of limited financial abilities.

5) The scheme will also provide enhanced security to the confidential data having huge importance, and hence can be used in places, where handling of sensitive data of high priority takes place

VIII. ACKNOWLEDGEMENT

We would like to take this opportunity to Show our gratitude to the faculty of the Electronics and Communication Department of our college, including our Head of the Department, Prof. Sudip Dogra who provided us with invaluable contributions regarding our present work. This achievement is also dedicated to our Administrator Mr. Satyen Mitra, who provided continuous support for this work. A special mention is made here about our friend Ms. Emon Dastider, who helped us with the composition of our document. And finally, we would like to thank Prof. Subir Kr. Sarkar for guiding us through this project.

IX. CONCLUSION

The basic advantage of our scheme lies in its cost-effectiveness, and its simple design. Once it is implemented on a commercial basis, it will establish itself as a great hindrance to the degraded practice of Software-piracy. There is also scope for more development in the design, which will enhance its efficiency and security

REFERENCES

[1] "RFID handbook: applications, technology, security, and privacy" by Syed Ahson and Mohammad Ilyas. CRC Press , Boca Raton
[2] "RFID Technology & Applications" by Stephen B. Miles, Sanjay E. Sharma & John R. Williams. Cambridge University Press, New York.
[3] G. O. Young, "Synthetic structure of industrial plastics (Book style with paper title and editor),"     in *Plastics*, 2nd ed. vol. 3, J. Peters, Ed. New York: McGraw-Hill, 1964, pp. 15–64.
[4] The Effect of Piracy on Markets for Consumer Transmutation Rights Lang, K.R.; Shang, R.D.; Vragov, R.; System Sciences, 2009. HICSS '09. 42nd Hawaii International Conference on 5-8 Jan. 2009
[5] Method based static software birthmarks: A new approach to derogate software piracy Mahmood, Y.; Sarwar, S.; Pervez, Z.; Ahmed, H.F.; Computer, Control and Communication, 2009. IC4 2009. 2nd International Conference on 17-18 Feb. 2009
[6] An intention model-based study of software piracy Tung-Ching Lin; Meng Hsiang Hsu; Feng-Yang Kuo; Pei-Cheng Sun; System Sciences, 1999. HICSS-32. Proceedings of the 32nd Annual Hawaii International Conference on Volume Track5,  5-8 Jan. 1999
[7] Understanding the behavioral intention to digital piracy in virtual communities - a propose model Kwong, T.C.H.; Lee, M.K.O.; e-Technology, e-Commerce and e-Service, 2004. EEE '04. 2004 IEEE International Conference on 28-31 March 2004

AUTHORS PROFILE

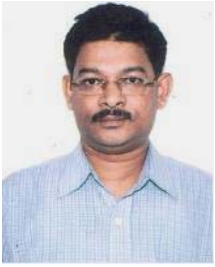

**Sudip Dogra**: Sudip Dogra received the B. Tech and M. Tech. Degree from the Institute of Radio Physics and Electronics, University of Calcutta in 1996 and 2003, respectively. He is doing PhD at Jadavpur University. He served Andrew Yule & Company Limited (A Govt. Of India Enterprise) as a Development Engineer( R & D Dept.) for about 6 years before coming to teaching profession. He joined as a faculty member in the Dept. of Electronics and Communication Engineering, Meghnad Saha Institute of Technology, Kolkata in 2003. Presently he is Assistant Professor and Head of the Department in Electronics & Communication Engineering Department of Meghnad Saha Institute of Technology, Kolkata. He has published more than 25 technical research papers in journals and peer – reviewed conferences. His most recent research focus is in the areas of 4th Generation Mobile Communication, MIMO, OFDM, WiMax, UWB, RFID& its applications etc.

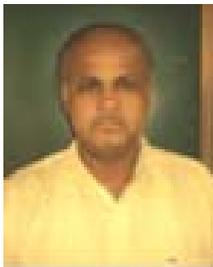

**Subir Kumar Sarkar** completed his B. Tech and M. Tech. from Institute of Radio Physics and Electronics, University of Calcutta in 1981 and 1983, respectively. He was in industry for about 10 years before coming to teaching profession.

He completed his Ph.D. (Tech) Degree from University of Calcutta in Microelectronics. Currently he is a professor in the Department Electronics and telecommunication Engineering, Jadavpur University. His present field of interest includes nano, single electron and spintronic device based circuit modeling, wireless mobile communication and data security in computer networks.

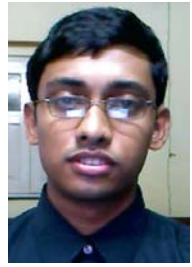

**Saustav Ghosh** is pursuing his Bachelor's degree in Electronics & Communication Engineering in Meghnad Saha Institute of Technology. He has published more than 6 technical research papers in journals and peer reviewed national and International Conferences. His earlier works were done in the fields of 4G Mobile communications, Co-operation in Mobile Communication, Mobile Security and WiMAX. His present field of interest is RFID and it's application.

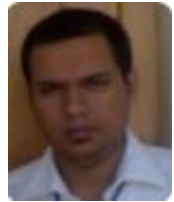

**Ritwik Ray** is pursuing his Bachelor's degree in Electronics & Communication Engineering in Meghnad Saha Institute of Technology. He has published more than 6 technical research papers in journals and peer reviewed national and International Conferences. His earlier works were done in the fields of 4G Mobile communications, Co-operation in Mobile Communication, Mobile Security and WiMAX. His present field of interest is RFID and it's application.

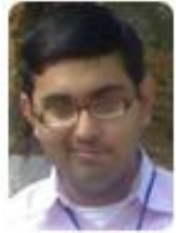

**Debharshi Bhattacharya** is pursuing his Bachelor's degree in Electronics & Communication Engineering in Meghnad Saha Institute of Technology. He has published more than 6 technical research papers in journals and peer reviewed national and International Conferences. His earlier works were done in the fields of 4G Mobile communications, Co-operation in Mobile Communication, Mobile Security and WiMAX. His present field of interest is RFID and it's application.